# A New Model to Analyze Power and Communication System Intra-and-Inter Dependencies

Sohini Roy[1], Harish Chandrasekaran[2], Anamitra Pal[2], Arunabha Sen[1]
[1]School of Computing, Informatics and Decision System Engineering
[2]School of Electrical, Computer and Energy Engineering
Arizona State University
Tempe-85281, Arizona, USA
Email: {sohini.roy, hchandr5, anamitra.pal, asen}@asu.edu

*Abstract*—The reliable and resilient operation of the smart grid necessitates a clear understanding of the intra-and-inter dependencies of its power and communication systems. This understanding can only be achieved by accurately depicting the interactions between the different components of these two systems. This paper presents a model, called modified implicative interdependency model (MIIM), for capturing these interactions. Data obtained from a power utility in the U.S. Southwest is used to ensure the validity of the model. The performance of the model for a specific power system application namely, state estimation, is demonstrated using the IEEE 118-bus system. The results indicate that the proposed model is more accurate than its predecessor, the implicative interdependency model (IIM) [1], in predicting the system state in case of failures in the power and/or communication systems.

## I. INTRODUCTION

Maintaining a sustainable lifestyle is contingent upon an uninterrupted supply of electricity. Modern power utilities try to ensure the continuity of this supply by operating an intricate network that consists of intra-and-inter-dependent power and communication system entities. For example, the power network measurements of the smart grid obtained by its sensors must be transferred to the control center by the communication entities. At the same time, the communication network entities themselves need power from the smart grid for their continued functionality. This interdependency has become critical in a smart grid environment where the failure of an entity in one network can lead to failures of the entities of the other network. Thus, it is essential to understand the interdependencies between the two types of networks for predicting the effect of failure of one or more entities on the overall system state. An inaccurate prediction may impact the decision making of an operator which can then lead to a less efficient operation of the grid.

Models that have been proposed previously to describe the intra-and-inter dependencies of critical infrastructures (such as [2],[3]) often lack physical realism as they are too simple to correctly portray the complex structure of the interdependent networks [4]. A specific drawback pertaining to the electrical infrastructure is the lack of clarity in the description of its communication network design. For example, in [5] a design of the joint network was given for the IEEE 14-bus system. However, the details of the information and communication technology (ICT) network were missing. The implicative interdependency model (IIM) [1] was successful in representing the complex interdependencies of a joint network using simple yet accurate Boolean logic-based Inter-Dependency Relations (IDRs). However, it also failed to accurately model the communication network entities as it lacked knowledge of the communication network design. With the help of a power utility in the U.S. Southwest, this paper presents a realistic design of the structure and operation of the power-and-communication network of a typical smart grid.

The rest of the paper is structured as follows. Section II of this paper explains our design principles by superimposing a synthetic yet realistic communication network on to a power network. Section III summarizes the differences between IIM and the proposed modified IIM (MIIM) using the concept of Inter-Dependency Relations (IDRs). A case study using the IEEE 14-bus is presented in Section IV to explain how the two models perform when a failure occurs in the system. Section V demonstrates the performance of IIM and MIIM on state estimation using the IEEE 118-bus system. Section VI concludes the paper and provides the scope for future work.

## II. DESIGNING OF A REALISTIC JOINT NETWORK

In this paper, the smart grid is viewed as a multilayer network, where entities in power layer (layer 1) are called $P$ type entities, $P = \{P_1, P_2, ... P_m\}$, entities in communication layer (layer 2) are called $C$ type entities, $C = \{C_1, C_2, ... C_n\}$, and entities which belong to both the layers (layer 3) are called $CP$ type entities, $CP = \{CP_1, CP_2, ... CP_o\}$. Fig. 1. classifies the joint network entities into these three categories. The figure also provides subdivisions of each of the three types of entities and the nomenclature assigned to them.

In Fig. 1, the $P$ type entities are subdivided into buses, transmission lines/transformers, and battery backup. The $C$ type entities are subclassified as substation entities (Type 1), synchronous optical networking (SONET)-ring entities (Type 2), or dense wavelength division multiplexing (DWDM)-ring entities (Type 3). Subdivisions of each of the three types of $C$ type entities are also shown in Fig. 1. The $CP$ type entities consist of $L$ type entities (power supply channels to different $C$ type entities), $R$ type entities (corresponding to remote

This research was supported in part by the NSF Grant EFMA award number 1441214.

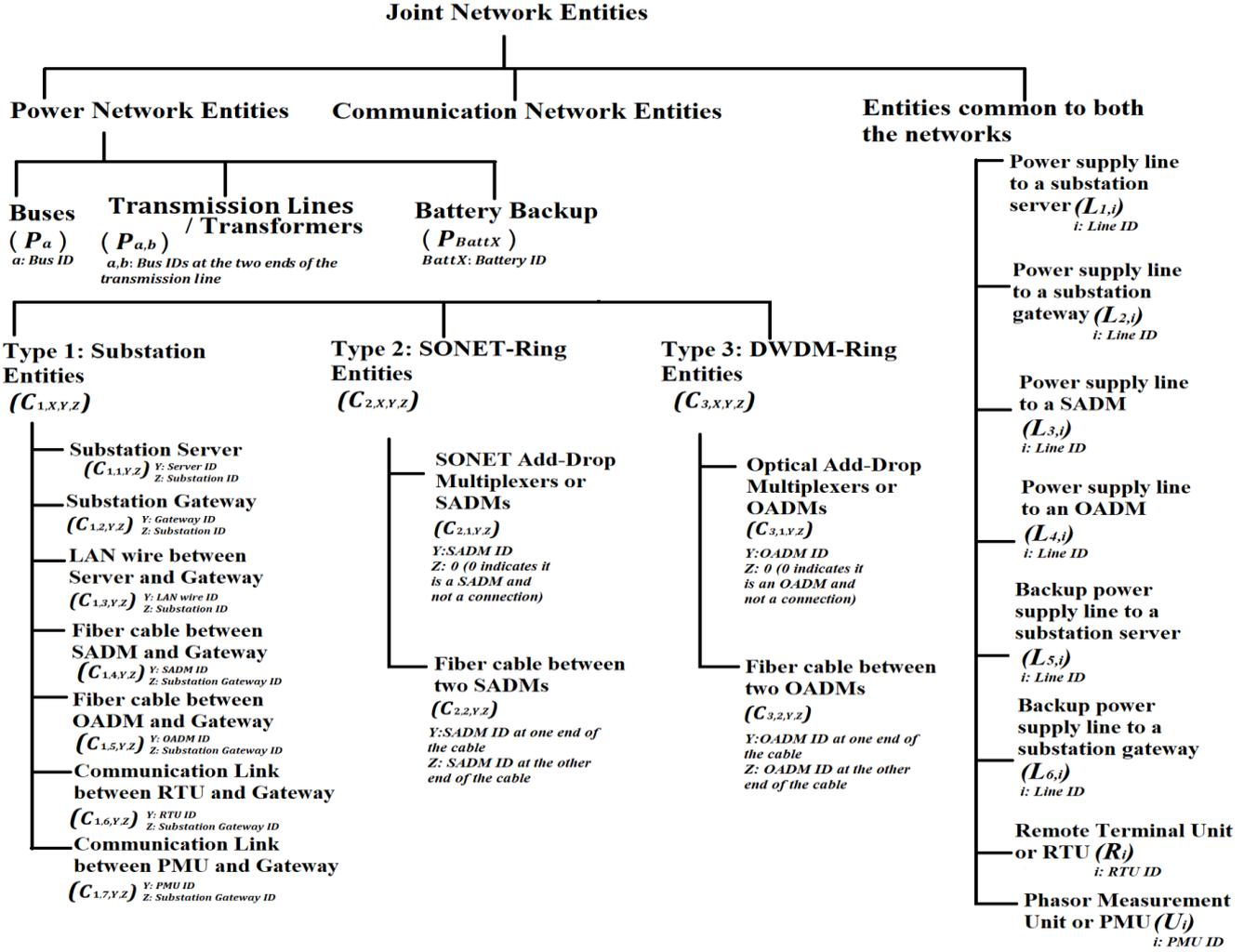

Fig. 1. Nomenclature for every entity of the joint network

terminal units (RTUs)), or $U$ type entities (corresponding to phasor measurement units (PMUs)). The design principles of the joint network are explained below in more details.

*A. Grouping buses into substations*

The buses of the power network are grouped into substations based on the logic given in [6]. The substation specific communication entities are Type 1 entities of Fig. 1. This step is further subdivided into the following sub-steps:

*1) Placing substation servers and gateways:* The substation server $(C_{1,1,Y,Z})$ is the main computing device of a substation. The supervisory control and data acquisition (SCADA) system inputs from RTUs and the synchrophasor system inputs from PMUs, reach the substation server via the gateway $(C_{1,2,Y,Z})$. The substation servers of the control centers use SCADA/PMU data to perform state estimation. Substation servers of other substations compress and encrypt SCADA/PMU data to forward them to the substation gateway. The gateway then sends the SCADA and PMU data to the control centers through the low bandwidth optical channels using SONET over Ethernet (SONEToE) [7] and high bandwidth optical channels using Ethernet over DWDM (EoDWDM) [8], respectively. Hence, the gateway connects the substation server to the rest of the communication network outside the substation and also to the PMUs and RTUs within the substation. Any data coming to and going from the substation server must pass through the gateway. The gateway also has a firewall that protects the server from cyber-attacks. The server is connected to the gateway via LAN connection $(C_{1,3,Y,Z})$.

*2) Supplying power to the Type 1 ICT entities:* The substation server and gateway receive power from the buses inside the substation. In order to avoid power outage within the substation, a battery backup $(P_{BattX})$ is also present in every substation. The battery supplies power to the Type 1 ICT entities when the buses in the substation do not have power.

*3) Placing two geographically diverse fiber optic cables from each substation:* There are two types of fiber optic channels going out from the gateway of each substation. One is the low bandwidth cable $(C_{1,4,Y,Z})$ that uses SONEToE technology and the other is the high bandwidth cable $(C_{1,5,Y,Z})$ which uses EoDWDM technology. In order to observe the performance of the synthetic network under different scenarios, two different cases are considered in this paper with respect to data transfer via the optical fiber cables.

In Case 1, the SONEToE channels are responsible for carrying RTU data to the nearest SONET-add-drop-multiplexer (SADM) of the SONET-ring (elaborated in Step 4 of this section) while the high bandwidth EoDWDM channel can only carry PMU data to the nearest optical-add-drop-multiplexer (OADM) of the DWDM-ring (elaborated in Step 5 of this section). In Case 2, under normal conditions, the low bandwidth SONEToE cable is responsible for carrying the RTU data to the nearest SADM of the SONET-ring while the high bandwidth channel is responsible for carrying the faster PMU data to the OADM of the DWDM-ring. However, in case of failure of the low bandwidth channel, in Case 2 (unlike Case 1), the EoDWDM channel can transmit SCADA inputs from the gateways to the SADMs. For fault tolerance, the control center gateways are connected to every SADM in the SONET-ring via multiple low bandwidth channels and also to every OADM in the DWDM-ring via multiple high bandwidth channels. As an illustration, Fig. 2 shows the substation division of the IEEE 14-bus system along with the substation servers and gateways. The $(C_{1,4,Y,Z})$ and $(C_{1,5,Y,Z})$ cables are shown in Fig. 3 and Fig. 4, respectively.

  *4) Placing RTUs and PMUs:* Every substation has RTUs ($R_i$). However, due to budget constraints, PMUs ($U_i$) are placed in only some of the substations using the methodology proposed in [9]. $R_i$ and $U_i$ measure SCADA system input data and synchrophasor system input data from the buses inside the substation and send them to the substation gateway via communication channels $(C_{1,6,Y,Z})$ and $(C_{1,7,Y,Z})$, respectively.

## B. Step 2: Finding shortest distance between all pairs of substations and selection of control centers

In this step, the distance between a connected pair of substations is calculated first based on the length of the transmission line connecting them. The distance between all pairs of substations is calculated next using the Floyd Warshall's all-pair shortest path algorithm [10]. Finally, two of the substations that are centrally located in the network and have large number of outgoing connections are selected as the primary and back-up control centers, respectively. For the IEEE 14-bus system, substation 2 is selected as the primary control center and substation 1 is selected as the secondary or backup control center (see Fig. 2). Note that this step is needed for the realistic placement of the SONET and DWDM-Rings in a synthetic system, as elaborated in the subsequent steps. This step can be skipped if the locations of the SADMs, OADMs, and control centers are known in advance.

## C. Step 3: Placement of SADMs and formation of SONET-Ring

SONEToE is a popular communication technology in which the SONET frames are directly carried on the Ethernet link layer. In this paper, SONEToE technology [7] is used for transmitting RTU data from the substations to the control centers. SADMs are located in close proximity to the generating substations and the control centers as they are the most important substations of the system. Other substations transmit their SCADA data to the nearest SADM using the low bandwidth Ethernet channels. For fault tolerance all such SADMs are connected to each other via a ring structure, termed SONET-Ring. The link between two SADMs is bi-directional; therefore, even if a single link or node in the ring fails, the ring as a whole continues to function normally. In normal conditions, data from every SADM is sent to the control centers directly. However, if a link between a control center and an SADM fails, data from that SADM is forwarded to the next SADM in the ring which in turn forwards the data to the control centers.

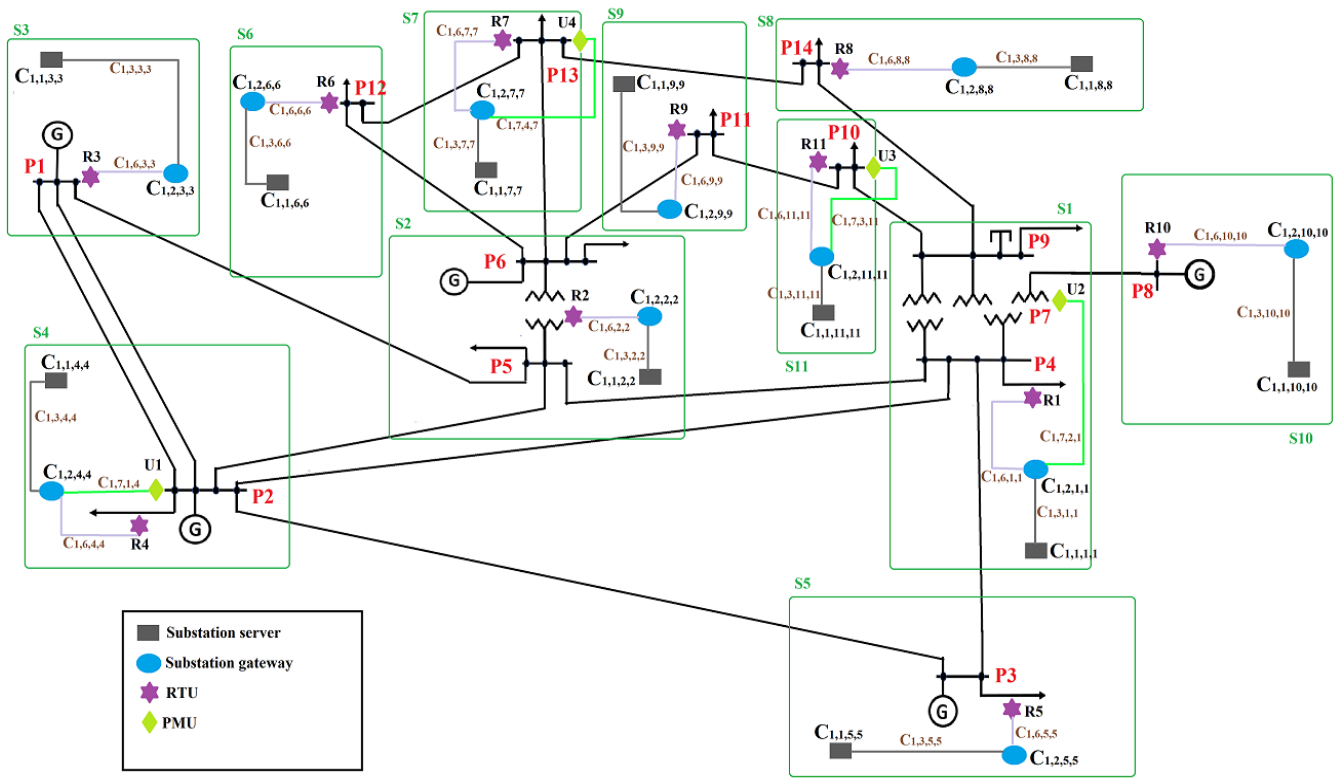

Fig. 2. Substation entities and substation division of IEEE 14-bus system

In case of the IEEE 14-bus system, SADMs are placed near S2 (main control center), S1 (back up control center) and S3, S4, S5, and S10 (generating substations). Therefore, a total of six SADMs are placed in this system (see Fig. 3). Gateways of all other substations are connected to the nearest SADM in the network. Each substation is thus connected to an SADM, except, the control centers which are connected to all the SADMs in the ring. Fig. 3 shows the SONET-Ring structure of the IEEE 14-bus joint network. The control centers are placed in the center of the ring to show the star-ring topology of the network.

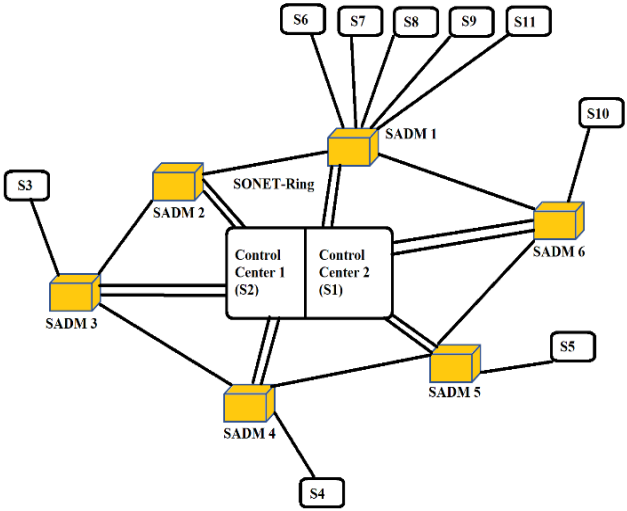

Fig. 3. SONET-Ring structure of IEEE 14-Bus system

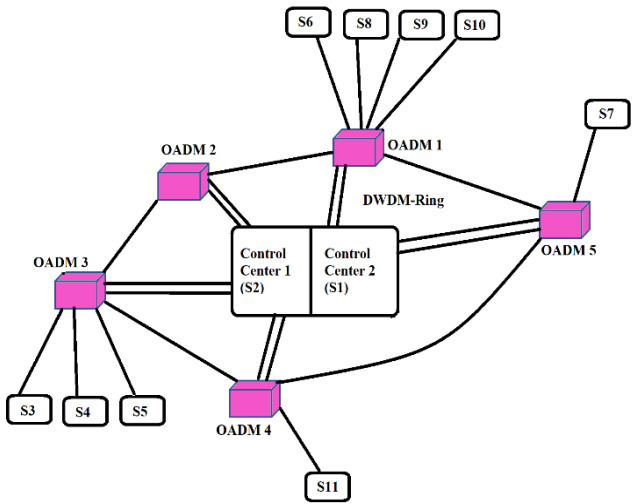

Fig. 4. DWDM-Ring structure of IEEE 14-Bus system

### D. Step 4: Placement of OADMs and formation of DWDM-Ring

EoDWDM [8] is a low-cost high bandwidth technology that automates network management for better scalability and performance. This technology can automatically detect problems across the entire network and resolve them very fast. The proposed synthetic network uses this technology for the transfer of high-speed PMU data from the substations that have PMUs, to the control centers. Note that not all substations currently have PMUs. However, considering the steady growth in the field of ICT and the popularity of PMUs, the proposed design assumes that PMUs will be placed in every substation in the near future. Therefore, by default, every substation has a high bandwidth EoDWDM channel coming out of it and ending in the DWDM-Ring.

The DWDM-Ring of the proposed design is composed of low cost OADMs, each of which is placed near a substation having a PMU inside it. An OADM is placed near each of the control centers irrespective of whether it contains a PMU or not. Similar to the SONET-Ring, the DWDM-Ring is also bi-directional, ensuring fault tolerance. In case of the IEEE 14-bus system, an OADM is placed near S2 (main control center), S1 (back up control center) and S4, S7, and S11 (PMU installed substations). Therefore, a total of five OADMs are placed in this system. Fig. 4 shows the DWDM-Ring structure of the IEEE 14-bus joint network.

### III. OVERVIEW OF MIIM AND MODELING OF IDRs

IDRs are logical equations that capture the interdependencies between two interacting entities. If they are written correctly, then by simply solving the IDRs after an entity or a set of entities have failed, the entities that will fail next can be identified. In this section, we describe creation of IDRs for the smart grid using MIIM. A smart grid system can be represented by the set $J(E, F(E))$, where $E$ is the set of entities in the joint network belonging to layers 1, 2, and 3 ($E = P \cup C \cup CP$) and $F(E)$ is the set of IDRs. In IIM [1], which was a precursor to MIIM, only structural dependencies were considered to formulate the IDRs. However, MIIM IDRs considers both the structural as well as the operational aspects of each of the entities during its formulation. In IIM, every entity was assigned a value of 0 or 1 depending on whether the entity was non-operational or operational. There was no concept of *reduced operability* in IIM, which is a common feature of most real systems. The entities in MIIM can take the following values: *0 indicating no operation*, *1 indicating reduced operation*, and *2 indicating full operation*.

From an implementation viewpoint, IIM IDRs were composed of two standard Boolean operations, namely, AND (denoted by '.') and OR (denoted by '+'). In contrast, MIIM uses three new Boolean operators for formulating the IDRs. The first operator is min-AND, denoted by '○', which selects the lowest of its input values. The second operator is max-OR, denoted by '●', which selects the highest of its input values. The third operator is new_XOR, which is denoted by '◉'. If all the inputs of new_XOR are same, then the output is also the same as the inputs; in all other cases the output is 1. The truth table for min-AND, max-OR, and new_XOR are given in Table I.

TABLE I. TRUTH TABLE FOR PROPOSED NEW OPERATIONS

| Input 1 | Input 2 | min-AND | max-OR | new_XOR |
|---------|---------|---------|--------|---------|
| 2 | 2 | 2 | 2 | 2 |
| 2 | 1 | 1 | 2 | 1 |
| 2 | 0 | 0 | 2 | 1 |
| 1 | 2 | 1 | 2 | 1 |
| 1 | 1 | 1 | 1 | 1 |
| 1 | 0 | 0 | 1 | 1 |
| 0 | 2 | 0 | 2 | 1 |
| 0 | 1 | 0 | 1 | 1 |
| 0 | 0 | 0 | 0 | 0 |

We now illustrate the process of creating IDRs using MIIM by deriving the IDRs for every entity of the IEEE 14-bus system. According to the design in Fig. 3, every SADM is connected to its neighboring SADMs in the ring. In the MIIM

IDR of $SADM_1$ ($C_{2,1,1,0}$) of the IEEE 14-bus system, this is expressed as:

$$C_{2,1,1,0} \leftarrow \underbrace{[(C_{2,1,2,0} \circ C_{2,2,1,2}) \bullet (C_{2,1,6,0} \circ C_{2,2,1,6})]}_{\equiv A} \quad (1)$$

Here, $C_{2,1,X,0}$ denotes $SADM_X$, while $C_{2,2,1,X}$ denotes the connection between $SADM_1$ and $SADM_X$, where X is the SADM ID. This IDR implies that $SADM_1$ remains operational if either the combination of $C_{2,1,2,0}$ AND $C_{2,2,1,2}$ is operational OR the combination of $C_{2,1,6,0}$ AND $C_{2,2,1,6}$ is operational. The SADMs can also forward the SCADA data collected from different substations to the control centers directly. Therefore, the MIIM IDR can be modified as:

$$C_{2,1,1,0} \leftarrow (A) \bullet \underbrace{[(C_{1,2,2,2} \circ C_{1,4,1,2}) \bullet (C_{1,2,1,1} \circ C_{1,4,1,1})]}_{\equiv B} \quad (2)$$

Here, $C_{1,2,X,X}$ is the gateway of control center X and $C_{1,4,1,X}$ is the connection between $SADM_1$ and the gateway of X. Now, this $SADM_1$ also depends on the gateways of substations 2, 6, 7, 8, 9, and 11 for collecting SCADA data (see Fig. 3). Therefore, the IDR is updated as follows:

$$C_{2,1,1,0} \leftarrow (B) \circ \big((C_{1,2,2,2} \circ C_{1,4,1,2}) \circledcirc (C_{1,2,6,6} \circ C_{1,4,1,6}) \circledcirc (C_{1,2,7,7} \circ C_{1,4,1,7}) \circledcirc (C_{1,2,8,8} \circ C_{1,4,1,8}) \circledcirc (C_{1,2,9,9} \circ C_{1,4,1,9}) \circledcirc (C_{1,2,11,11} \circ C_{1,4,1,11})\big) \equiv C \quad (3)$$

If all the gateways (2,6,7,8,9,11) from which $SADM_1$ receives SCADA data, remain operational then $SADM_1$ will work at its highest level of operation, i.e. 2. If one or more gateways fail or the connection between one such gateway and $SADM_1$ fails, then $SADM_1$ will work at a reduced level of operation, i.e. 1. If all the gateways connected to the SADM fails, then the SADM will also fail as it will have no data to carry to the control centers. Lastly, the SADM needs power supply to function. Hence, the IDR is further modified as:

$$SADM_1(C_{2,1,1,0}) \leftarrow (C) \circ [(P_4 \circ L_{3,1}) \bullet (P_7 \circ L_{3,2}) \bullet (P_9 \circ L_{3,3}) \bullet (P_5 \circ L_{3,4}) \bullet (P_6 \circ L_{3,5}) \bullet (P_{12} \circ L_{3,6}) \bullet (P_{13} \circ L_{3,7}) \bullet (P_{14} \circ L_{3,8}) \bullet (P_{11} \circ L_{3,9}) \bullet (P_{10} \circ L_{3,10})] \quad (4)$$

The above final IDR of $SADM_1$ implies that it can receive power supply from any of the buses of any of the substations it is connected to; $P_4, P_7, P_9, P_5, P_6, P_{12}, P_{13}, P_{14}, P_{11}, P_{10}$ and $L_{3,1}, L_{3,2}, L_{3,3}, L_{3,4}, L_{3,5}, L_{3,6}, L_{3,7}, L_{3,8}, L_{3,9}, L_{3,10}$ are the buses and power supply lines respectively to $SADM_1$. For the $SADM_1$ to work, it should receive power from at least one of these buses. In this manner, the IDRs for all the six SADMs in the IEEE 14-bus system can be formulated. For creating the corresponding IIM IDRs, the '$\circ$' and '$\circledcirc$' operators must be replaced by '.' and the '$\bullet$' operator must be replaced by '+'. Similarly, the IDRs of OADMs can also be formulated for both MIIM and IIM.

Now, the IDR for the gateway of substation 1 can be formulated using the following set of steps:

*Step 1:* The substation gateway depends on the RTU of that substation for receiving SCADA data. This is described by, $\quad C_{1,2,1,1} \leftarrow (R_1 \circ C_{1,6,1,1}) \equiv D \quad (5)$ where $C_{1,2,1,1}$ is the gateway of substation 1, $R_1$ is the RTU of that substation and $C_{1,6,1,1}$ is the communication channel connecting the RTU to the gateway. If a substation has multiple RTUs, then the gateway of that substation collects data from all the RTUs of that substation.

*Step 2:* The substation gateway should also remain connected to at least one of the SADMs. It can receive SCADA data from other substations (if the gateway belongs to a control center) or it can send SCADA data to the control centers through the SONET-Ring. Also, if the gateway is connected to an SADM but the RTU of the substation does not work, then the gateway will not be able to send any data to the SADM. Finally, if one (or more in the case of control centers) SADM(s) connected to the gateway fail then the gateway will work at a reduced level of operation. This is described by the following IDR:

$$C_{1,2,1,1} \leftarrow [(D) \circ \big((C_{2,1,1,0} \circ C_{1,4,1,1}) \circledcirc (C_{2,1,2,0} \circ C_{1,4,2,1}) \circledcirc (C_{2,1,3,0} \circ C_{1,4,3,1}) \circledcirc (C_{2,1,4,0} \circ C_{1,4,4,1}) \circledcirc (C_{2,1,5,0} \circ C_{1,4,5,1}) \circledcirc (C_{2,1,6,0} \circ C_{1,4,6,1})\big)] \equiv E \quad (6)$$

In this IDR, $C_{2,1,X,0}$ is $SADM_X$ and $C_{1,4,X,Y}$ are the ethernet connections between $SADM_X$ and $Gateway_Y$.

*Step 3:* The substation gateway is also dependent on the PMU of that substation for receiving PMU data, i.e.

$$C_{1,2,1,1} \leftarrow (U_2 \circ C_{1,7,2,1}) \equiv F \quad (7)$$

In this IDR, $U_2$ is the PMU of that substation and $C_{1,7,2,1}$ is the communication channel connecting the PMU to the gateway. Similar to the case of RTUs, if a substation has multiple PMUs, then the gateway of that substation collects data from all the PMUs of that substation.

*Step 4:* The gateway should also remain connected to at least one OADM (similar to Step 2 in the case of SADMs). Hence,

$$C_{1,2,1,1} \leftarrow [(F) \circ \big((C_{3,1,1,0} \circ C_{1,5,1,1}) \circledcirc (C_{3,1,2,0} \circ C_{1,5,2,1}) \circledcirc (C_{3,1,3,0} \circ C_{1,5,3,1}) \circledcirc (C_{3,1,4,0} \circ C_{1,5,4,1}) \circledcirc (C_{3,1,5,0} \circ C_{1,5,5,1}) \circledcirc (C_{3,1,6,0} \circ C_{1,5,6,1})\big)] \equiv G \quad (8)$$

In the above IDR, $C_{3,1,X,0}$ is $OADM_X$ and $C_{1,5,X,Y}$ implies the DWDM connections between $OADM_X$ and $Gateway_Y$.

*Step 5:* The gateway should receive power from at least one of the buses in that substation.

*Step 6:* The gateway should remain connected to the substation server.

In order to obtain SCADA data from the buses of a substation, Steps 1, 2, 5, and 6 should be followed. Thus, the following IDR can be used to determine if a gateway is operational with respect to SCADA data. In other words, if the evaluation of the following IDR results in 2 (highest operational level) or 1 (reduced operational level), then the SCADA data from the corresponding buses can be received by the server of the substation.

$$Gateway_1^{SCADA}(C_{1,2,1,1}) \leftarrow [C_{1,1,1,1} \circ C_{1,3,1,1}] \circ [E] \circ [(P_4 \circ L_{2,4}) \bullet (P_7 \circ L_{2,7}) \bullet (P_9 \circ L_{2,9}) \bullet (P_{Batt1} \circ L_{6,1})] \quad (9)$$

The above IDR is the final $Gateway_1^{SCADA}(C_{1,2,1,1})$ IDR for Case 1 where strictly separate channels are used for RTU and PMU data. However, for Case 2, if all the connections to the SADMs fail, the gateway can still receive SCADA data from the other substations if the data is sent through the high bandwidth EoDWDM network, i.e. through the OADMs. Therefore, the above IDR can be further modified for Case 2 as shown below:

$$Gateway_1^{SCADA}(C_{1,2,1,1}) \leftarrow [C_{1,1,1,1} \circ C_{1,3,1,1}] \circ [(D) \circ$$
$$(((C_{2,1,1,0} \circ C_{1,4,1,1}) \odot (C_{2,1,2,0} \circ C_{1,4,2,1}) \odot (C_{2,1,3,0} \circ C_{1,4,3,1}) \odot (C_{2,1,4,0} \circ C_{1,4,4,1}) \odot (C_{2,1,5,0} \circ C_{1,4,5,1}) \odot (C_{2,1,6,0} \circ C_{1,4,6,1})) \bullet ((C_{3,1,1,0} \circ C_{1,5,1,1}) \odot (C_{3,1,2,0} \circ C_{1,5,2,1}) \odot (C_{3,1,3,0} \circ C_{1,5,3,1}) \odot (C_{3,1,4,0} \circ C_{1,5,4,1}) \odot (C_{3,1,5,0} \circ C_{1,5,5,1}) \odot (C_{3,1,6,0} \circ C_{1,5,6,1})))] \circ [(P_4 \circ L_{2,4}) \bullet (P_7 \circ L_{2,7}) \bullet (P_9 \circ L_{2,9}) \bullet (P_{Batt1} \circ L_{6,1})] \quad (10)$$

In order to obtain PMU data from the buses of a substation, Steps 3, 4, 5, and 6 should be followed. The following IDR can be used to determine if a gateway is operational with respect to PMU data. In other words, if the evaluation of the following IDR results in 2 (highest operational level) or 1 (reduced operational level), then the PMU data from the corresponding buses can be received by the server of the substation.

$$Gateway_1^{PMU}(C_{1,2,1,1}) \leftarrow [C_{1,1,1,1} \circ C_{1,3,1,1}] \circ [G] \circ [(P_4 \circ L_{2,4}) \bullet (P_7 \circ L_{2,7}) \bullet (P_9 \circ L_{2,9}) \bullet (P_{Batt1} \circ L_{6,1})] \quad (11)$$

Now, gateway 1 is said to be fully operational if the following IDR gives a value 2, which implies that both SCADA and PMU data is sent (or received in case of control centers) by the gateway. If the IDR gives a value of 1, then it can be stated that either the PMU data or the SCADA data is sent/received by the gateway. If none of the two types of data is sent or received, then the evaluation of the following IDR will give 0.

$$C_{1,2,1,1} \leftarrow Gateway_1^{SCADA} \odot Gateway_1^{PMU} \quad (12)$$

A substation server depends on the substation gateway and the power supply links from at least one of the buses of the substation. Therefore, the IDR of the server of substation 1 of IEEE 14-bus system can be written as:

$$C_{1,1,1,1} \leftarrow (C_{1,2,1,1} \circ C_{1,3,1,1}) \circ [(P_4 \circ L_{1,4}) \bullet (P_7 \circ L_{1,7}) \bullet (P_9 \circ L_{1,9}) \bullet (P_{Batt1} \circ L_{5,1})] \quad (13)$$

In the above IDR, $L_{1,4}, L_{1,7}, L_{1,9}$ are the power supply lines to the server from buses $P_4, P_7, P_9$, respectively. $L_{5,1}$ is the power supply line to the gateway from the battery backup $P_{Batt1}$. Following these steps, the IDRs of the substation servers and substation gateways for every substation can be derived for a synthetic joint network of a power system.

## IV. CASE STUDY

The IEEE 14-bus network is used to illustrate the cascading failures that take place after a single failure occurs in the joint network. The failure which is simulated is a terrorist attack on substation 6 of this system. The physical attack leads to the immediate failure of Bus 12 ($P_{12}$), substation server ($C_{1,1,6,6}$), and substation gateway ($C_{1,2,6,6}$). The division of buses into substations for the IEEE 14-bus system is shown in Fig. 2 The operational statuses of the communication entities which transfer the data to the control center are calculated using MIIM IDRs and IIM IDRs, respectively. Table II shows how the smart grid system is affected gradually at each time step, denoted by $Ti$, if MIIM IDRs are employed. From Table II, it is observed that as a result of substation 6 failure, bus $P_{12}$, gateway and server inside substation 6 fails immediately (at time instant T1). Consequently, the SONEToE and EoDWDM channels coming out of gateway 6 fail at the next time instant (T2). At T3, $SADM_1$ and $OADM_1$ start working at a reduced level of operation as they are not getting the expected data from gateway 6, but still get data from the other gateways to which they are connected. The cascading failure of entities stops at T3. The results obtained using MIIM IDRs are same irrespective of whether data transmission is done on the basis of Case 1 or Case 2.

TABLE II.  FAILURE OF ENTITIES WITH TIME OBTAINED USING MIIM

| T1 | $P_{12} \to 0$ | $C_{1,1,6,6} \to 0$ | $C_{1,2,6,6} \to 0$ | | | | |
|---|---|---|---|---|---|---|---|
| T2 | $P_{12} \to 0$ | $C_{1,1,6,6} \to 0$ | $C_{1,2,6,6} \to 0$ | $C_{1,4,1,6} \to 0$ | $C_{1,5,1,6} \to 0$ | | |
| T3 | $P_{12} \to 0$ | $C_{1,1,6,6} \to 0$ | $C_{1,2,6,6} \to 0$ | $C_{1,4,1,6} \to 0$ | $C_{1,5,1,6} \to 0$ | $C_{2,1,1,0} \to 1$ | $C_{3,1,1,0} \to 1$ |

When IIM IDRs are used, two different results are obtained for the two cases of data transmission. Table III shows the cascading failure of entities obtained using IIM IDRs. The failure of entities at time instants T1 and T2 happen in the same way as in the case of MIIM. At T3, $SADM_1$ and $OADM_1$ fail completely due to the failure of one of the gateways (gateway 6) connected to them. This happens due to the binary nature of IIM, which does not account for reduced operability. Consequently, at time instant T4, no SCADA data is obtained from $P_{10}, P_{11}, P_{13}$, and $P_{14}$ for Case 1, and $P_{11}$ and $P_{14}$ for Case 2. More entities fail in Case 1 than in Case 2 because in Case 2, unlike Case 1, the high bandwidth channel is capable of carrying both RTU and PMU data (see Section II.A).

TABLE III.  FAILURE OF ENTITIES WITH TIME OBTAINED USING IIM

| T1 | $P_{12} \to 0$ | $C_{1,1,6,6} \to 0$ | $C_{1,2,6,6} \to 0$ | | | | | |
|---|---|---|---|---|---|---|---|---|
| T2 | $P_{12} \to 0$ | $C_{1,1,6,6} \to 0$ | $C_{1,2,6,6} \to 0$ | $C_{1,4,1,6} \to 0$ | $C_{1,5,1,6} \to 0$ | | | |
| T3 | $P_{12} \to 0$ | $C_{1,1,6,6} \to 0$ | $C_{1,2,6,6} \to 0$ | $C_{1,4,1,6} \to 0$ | $C_{1,5,1,6} \to 0$ | $C_{2,1,1,0} \to 0$ | $C_{3,1,1,0} \to 0$ | |
| T4 CASE 1 | $P_{12} \to 0$ | $C_{1,1,6,6} \to 0$ | $C_{1,2,6,6} \to 0$ | $C_{1,4,1,6} \to 0$ | $C_{1,5,1,6} \to 0$ | $C_{2,1,1,0} \to 0$ | $C_{3,1,1,0} \to 0$ | NO SCADA FROM $P_{10}, P_{11}, P_{13}, P_{14}$ |
| T4 CASE 2 | $P_{12} \to 0$ | $C_{1,1,6,6} \to 0$ | $C_{1,2,6,6} \to 0$ | $C_{1,4,1,6} \to 0$ | $C_{1,5,1,6} \to 0$ | $C_{2,1,1,0} \to 0$ | $C_{3,1,1,0} \to 0$ | NO SCADA FROM $P_{11}, P_{14}$ |

## V. STATE ESTIMATION RESULTS

State estimation is performed for the IEEE 118-bus system to (1) understand if MIIM can predict the system state more accurately than IIM, and (2) demonstrate the scalability of MIIM. The state estimation is performed considering a single entity or multiple entity failures and the states predicted using MIIM and IIM are both compared for the two different cases of communication discussed earlier.

## A. Overview of state estimation

The voltage magnitudes and angles (or the real and imaginary components of voltages) of all the buses constitute the states of the system. They are estimated using the formulated IDRs and the measurements obtained from the RTUs and PMUs. Note that loss of measurements from the sensors of a bus can result in a bad estimate of the state of that bus and/or the states of the neighboring buses. The relationship between the state matrix V and the measurement matrix Z for a bus that has a PMU placed on it, is given by:

$$Z = \begin{bmatrix} Z_r^S \\ Z_i^S \\ Z_r^P \\ Z_i^P \\ I_r \\ I_i \end{bmatrix} = \begin{bmatrix} 1 & 0 \\ 0 & 1 \\ 1 & 0 \\ 0 & 1 \\ C_1 & C_2 \\ C_3 & C_4 \end{bmatrix} \begin{bmatrix} V_r \\ V_i \end{bmatrix} \equiv JV \quad (14)$$

where $Z_r^S$, $Z_i^S$ denote the real and imaginary voltages estimated using the traditional SCADA-based state estimation [11], $Z_r^P$, $Z_i^P$ denote the real and imaginary voltage measurements obtained from the PMU, and $I_r$, $I_i$ denote the real and imaginary (branch) current measurements obtained from the PMU. The matrices $C_1, C_2, C_3, C_4$ which relate the (branch) current measurements to the states of the system are obtained from the admittance matrix of the system. For instance, if a branch 'ab' with a series impedance $g_{ab} + j\, b_{ab}$ and shunt admittance $g_{a0} + j\, b_{a0}$ has a current $I_{ab}$ flowing through it, then the relationship between the current in rectangular coordinates and the states of the system are given by:

$$\begin{bmatrix} (I_{ab})_r \\ (I_{ab})_i \end{bmatrix}$$
$$= \begin{bmatrix} g_{ab} & -g_{ab} & (-b_{ab} - b_{a0}) & b_{ab} \\ (b_{ab} + b_{a0}) & -b_{ab} & g_{ab} & -g_{ab} \end{bmatrix} \begin{bmatrix} (V_a)_r \\ (V_a)_i \\ (V_b)_r \\ (V_b)_i \end{bmatrix}$$
(15)

The matrix $J$ represents the matrix relating the measurements and the states of the system. $V_r, V_i$ denotes the real and imaginary estimates of the states. In (15), the relation between the measurements and the states is linear, which means that it can be solved using the weighted least squares approach:

$$V = (J^T W^{-1} J)^{-1} (J^T W^{-1}) Z \quad (16)$$

In (16), the matrix $W$ is the final weight matrix comprising error covariance matrices of both SCADA and PMU measurements in rectangular form. The matrix $W$ is obtained using the methodology developed in [12]. In this paper, the SCADA measurement errors are assumed to be from a Gaussian distribution with 0 mean and 3% standard deviation, while the errors in the PMU measurements are assumed to be from a Gaussian distribution with 0 mean and 0.1% standard deviation.

## B. Hardware failure of gateway 13 and SADM 39 of IEEE 118-bus system

The gateway 13 is connected to bus 13 and SADM 39 is placed at substation 76 containing bus 85. Failure of gateway 13 results in loss of SCADA measurements at bus 14 for MIIM. The same original event results in the loss of SCADA measurements at buses 12, 13, 14, and 16, and loss of PMU measurements at bus 12 for IIM. The above-mentioned failures are common to both Case 1 and Case 2. Subsequent case-specific results are described below.

Case 1: As the high bandwidth channel cannot be used for carrying both RTU data and PMU data in this case, it results in an additional loss of SCADA measurements at buses 84, 85, and 88 for both MIIM and IIM, and a loss of PMU measurement at bus 85 for IIM due to SADM 39 failure. The state estimation results are shown in Fig. 5, which depicts the absolute difference between the estimated value and the true value of the states for both the interdependency models. The buses 7, 12, 84, 85, 88, and 117 observe a significant difference between the estimated states for both the models. This is because, the buses 84 and 88 (7 and 117) are neighbors of bus 85 (12) which loses PMU data in the case of IIM, but not in the case of MIIM.

Case 2: In this case, the high bandwidth channel is capable of carrying both RTU and PMU data. Because of this, no subsequent failures take place for both IIM and MIIM. The results obtained on performing state estimation are shown in Fig. 6. The difference between the estimated states for both the models is considerable at buses 7, 12, and 117, due to the same reason mentioned in Case 1.

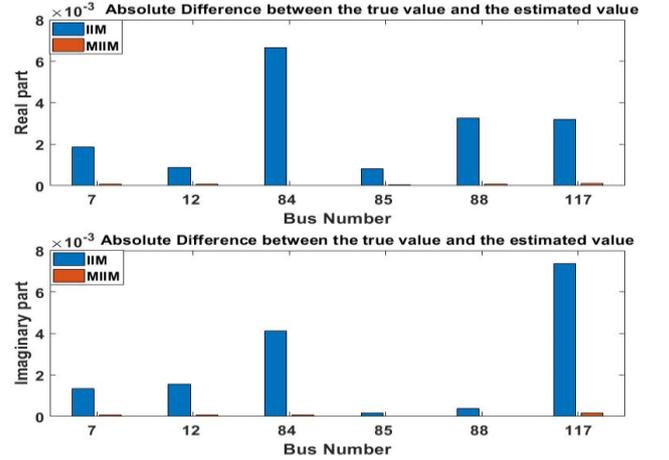

Fig. 5. State estimation result for gateway 13 and SADM 39 failure for Case 1

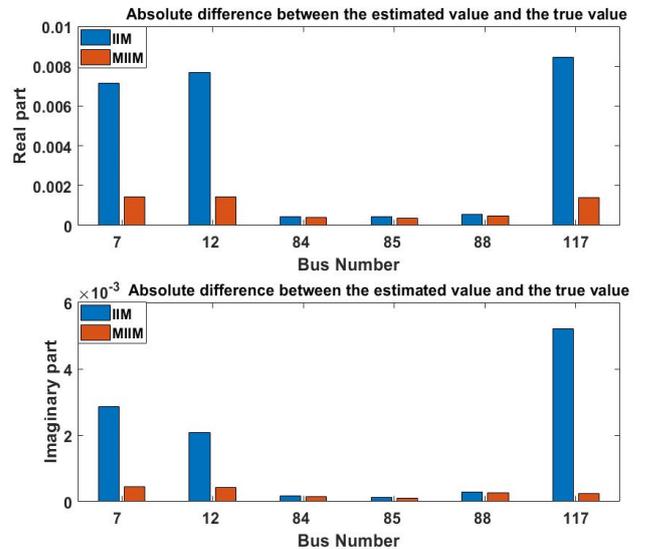

Fig. 6. State estimation result for gateway 13 and SADM 39 failure for Case 2

## C. Damage of substation 85 of IEEE 118-bus system

Substation 85 consists of bus 95 of the IEEE 118-bus system. Damage to this substation would result in loss of all communication entities placed at or connected to substation 85. This results in PMU measurement losses at bus 94 and SCADA measurement losses at bus 94, 95, and 100 for IIM. However, it results in measurement loss at only bus 95 for MIIM. The above-mentioned failures are common to both Case 1 and Case 2. Subsequent case-specific results are described below.

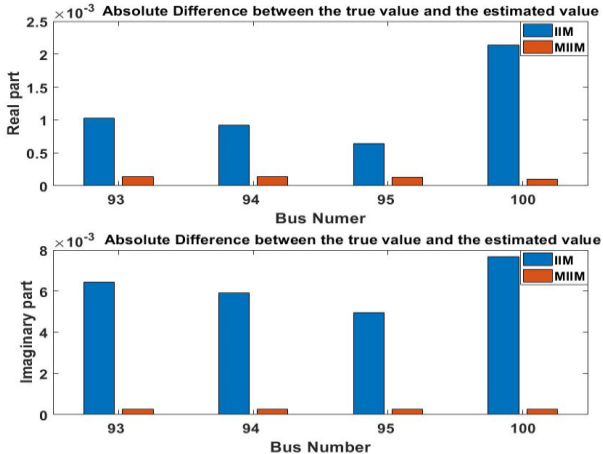

Fig. 7. State estimation result for substation 85 failure for Case 1

Case 2: In this case, since the high bandwidth channel is capable of carrying both RTU and PMU data, no subsequent failures take place for both IIM and MIIM. The results obtained on performing state estimation are shown in Fig. 8. The buses 93, 94, 95, and 100 observe a notable difference between the estimated states for both the models for the same reason mentioned in the previous case.

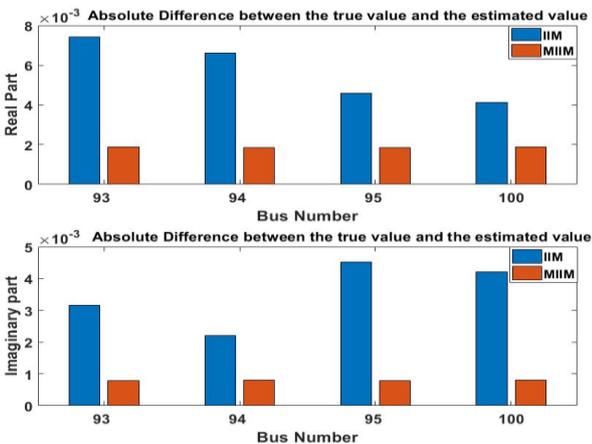

Fig. 8. State estimation result for substation 85 failure for Case 2

The results obtained above confirm that the states of the system estimated using MIIM are closer to the true values than the ones obtained using IIM.

## VI. CONCLUSIONS AND FUTURE WORK

An accurate estimation of the system's operating status in case of a failure, even before the failure occurs or even while that failure is manifesting, is extremely important and valuable. However, doing this for the modern smart grid is difficult because the intra-and-inter-dependencies between its power-and-communication networks are not well understood. The model presented in this paper, namely MIIM, is an effort by us to correctly capture these dependencies. The power system application that is used to demonstrate the practical utility of MIIM is state estimation. The results indicate that MIIM, in comparison to its predecessor, IIM, is more realistic in estimating the system state after some failure has occurred in the joint power-communication network. The future scope of work includes the use of MIIM for examining more complicated power/communication failure scenarios, such as EMP attacks, as well as for analyzing progressive recovery options of the joint network following a blackout/brownout.